\documentclass[runningheads,a4paper,oribibl]{llncs}

\usepackage{amsmath}
\usepackage{amssymb}
\usepackage{calc}
\usepackage[sort&compress,numbers,sectionbib]{natbib}

\usepackage{units}

\usepackage[a4paper, total={6in, 9in}]{geometry}
\usepackage[pdfencoding=auto, psdextra, hidelinks]{hyperref}
\usepackage{dcolumn}
	\newcolumntype{d}[1]{D{.}{.}{#1}}

\usepackage{booktabs}
\usepackage[font=footnotesize]{subcaption}

\usepackage[inline]{enumitem}

\usepackage{acronym}

\acrodef{arima}[ARIMA]{autoregressive integrated moving average}
\acrodef{ar}[AR]{autoregressive}
\acrodef{sarima}[SARIMA]{seasonal \ac{arima}}
\acrodef{nnr}[NNR]{nearest neighbor regression}
\acrodef{snnr}[SNNR]{seasonal \ac{nnr}}
\acrodef{knn}[kNN]{$k$ nearest neighbor}
\acrodef{svr}[SVR]{support vector regression}
\acrodef{ann}[ANN]{artificial neural networks}
\acrodef{rmse}[RMSE]{root mean square error}
\acrodef{nrel}[NREL]{National Renewable Energy Laboratory}
\acrodef{arm}[ARM]{Atmospheric Radiation Measurement}
\acrodef{ms}[MS]{model selection}
\acrodef{fc}[FC]{forecast}

\newcommand{\inNLargerZero}[1]{\ensuremath{#1 \in \mathbb{N}}}
\newcommand{\R}{\ensuremath{\mathbb{R}}}
\newcommand{\inNLargerEqualZero}[1]{\ensuremath{#1 \in \mathbb{N}_{0}}}
\newcommand{\tran}{\ensuremath{^{T}}} 

\usepackage{graphicx}

\begin{document}

\mainmatter

\title{Very Short Term Time-Series Forecasting of Solar Irradiance Without~Exogenous~Inputs\thanks{This work was partially supported by the German Federal Ministry for Economic Affairs and Energy (BMWi), Project No. 0324024A.}}

\author{Christian A. Hans \and Elin Klages\thanks{Currently at the Institute of Modelling and Computation, Technische Universit\"at Hamburg, Germany.}}
\institute{Control Systems Group, Technische Universit\"at Berlin, Germany}

\maketitle


\begin{abstract}
This paper compares different forecasting methods and models to predict average values of solar irradiance with a sampling time of \unit[15]{min} over a prediction horizon of up to \unit[3]{h}.
The methods considered only require historic solar irradiance values, the current time and geographical location, i.e., no exogenous inputs are used.
Nearest neighbor regression (NNR) and autoregressive integrated moving average (ARIMA) models are tested using different hyperparameters, e.g., the number of lags, or the size of the training data set, and data from different locations and seasons.
The hyperparameters and their effect on the forecast quality are analyzed to identify properties which are likely to lead to good forecasts.
Using these properties, a reduced search space is derived to identify good forecasting models much faster.
\end{abstract}



\section{Introduction}
\label{sec:introduction}


In the last years, the capacity of globally installed photovoltaic generators has continuously increased~\cite{Ren21}.
With this growth, the intermittent nature of their infeed has become a major challenge in the operation of electric grids \cite{Denholm2011}.
One important way to address this challenge is to use accurate forecasts to predict the infeed of photovoltaic power plants with a high temporal resolution \cite{West2014}.

Various studies on short term forecasting of
photovoltaic power plant infeed \cite{PedroCoimbra2012, PedroCoimbra2015, Shi2012, Bacher2009, QiaoZeng2012} and
solar irradiance \cite{Reikard2008, SfetsosCoonick2000, Mellit2010} have been published.
Autoregressive integrated moving average (ARIMA)\acused{arima} models have been widely adopted in this context \cite{Reikard2008, PedroCoimbra2012, Bacher2009}.
In the last decades forecasting methods that use techniques from artificial intelligence have become more prominent.
Most of them use \acl{ann} \cite{PedroCoimbra2012, QiaoZeng2012, Reikard2008, SfetsosCoonick2000, Mellit2010}.
Others employ support vector \cite{QiaoZeng2012, Shi2012} and nearest neighbor regression \cite{PedroCoimbra2012, PedroCoimbra2015}.

Despite additional effort that comes with the use of exogenous inputs, e.g., cloud cover or air temperature, only in \cite{PedroCoimbra2012, Bacher2009, SfetsosCoonick2000} forecasts without exogenous inputs are considered.
Moreover, in most publications \cite{PedroCoimbra2012, Bacher2009, Shi2012, SfetsosCoonick2000, Mellit2010, PedroCoimbra2015} the forecasting models are obtained using only one data set.
This makes it hard to draw general conclusions from the analysis that can be transferred to other locations or seasons of the year.
Furthermore, in \cite{PedroCoimbra2012, Mellit2010, Reikard2008, Shi2012, SfetsosCoonick2000, PedroCoimbra2015} the model selection process is not explicitly discussed.
Consequently, even though high forecast accuracies could be achieved for single data sets, due to the missing description of the selection processes, the findings cannot be directly used for different data sets.
For example, in \cite{Mellit2010} the authors state that they used a search and in \cite{Reikard2008} that they tried different model structures to choose an \acl{ann}.
However, they did not provide information on the search space.
In \cite{Shi2012}, a search space for \acl{svr} forecasts is provided but no information on the selection criterion or the final model structure and parameters is given.
The structure of an \acl{ann} forecasting model is examined in \cite{SfetsosCoonick2000} using a sensitivity analysis.
Unfortunately, the publication does not contain sufficient details to replicate the analysis for different data.
To the knowledge of the authors only in \cite{PedroCoimbra2015} a search space is provided and the results are analyzed to gain information about suitable model structures.
However, the analysis is only performed on data from one location.
This makes it hard to draw conclusions that allow identify a reduced search space and in practice often leads to an exhaustive search for suitable models.

In this paper we derive a significantly reduced search space that exhibits a high probability of finding an accurate forecasting model.
As potential forecasting methods, \ac{arima} and \ac{nnr} were considered due to their wide adoption.
Furthermore, their hyperparameters, e.g., the number of autoregressive lags or neighbors, are mostly discrete which allows to form a discrete search space where points in a certain range can be explored.
For each point, i.e., for each set of hyperparameters in this search space, forecasting models are trained and tested using data from different locations and seasons.
Based on the resulting forecast accuracies, conclusions for a reduced search space are drawn.
These include the choice of \ac{nnr} over \ac{arima},
the number of training data points and autoregressive lags
as well as
handling of night data and
use of transmissivity instead of irradiance.
The selection process is explained in much detail allowing others to perform a similar search on different data.

For simplicity and robustness, this work focuses on forecasts of solar irradiance without exogenous inputs using only historic irradiance data, time and location.
Motivated by the smallest intraday interval of energy trading in Germany a \unit[15]{min} prediction step is chosen \cite{EPEX2016}.
To cover a full charge or discharge of a medium size storage unit, a prediction horizon of \unit[3]{h} is~considered.

The remainder of this paper is organized as follows.
Solar irradiance and data preprocessing are discussed in Section~\ref{sec:solarRadiation}.
In Section~\ref{sec:timeSeriesForecasting}, basics on time-series forecasting are discussed.
The forecasting methods considered in this work are presented in Section~\ref{sec:forecastingMethods}.
Then, the model selection procedure is illustrated in Section~\ref{sec:modelSelection}.
Finally, in Section~\ref{sec:analysis} the results of the hyperparameters search are analyzed.

\subsection{Preliminaries}
Throughout this paper, real numbers are denoted by $\mathbb{R}$, nonnegative real numbers by $\mathbb{R}_{0}^+$, positive real numbers by $\mathbb{R}^+$, natural numbers by $\mathbb{N}$ and nonnegative integers by $\mathbb{N}_{0}$.
The transpose of a matrix $x$ is $x^T$ and the Euclidean norm of a vector $x$ is $\| x \|_2$.
The sum over all elements of a set $\mathbb{K}\subset\mathbb{N}$ where each element $i\in\mathbb{K}$ is taken exactly once is denoted $\sum_{i \in \mathbb{K}} x_i$.


\section{Solar Irradiance}
\label{sec:solarRadiation}
The sunlight reaching the outer earth's atmosphere is called extraterrestrial solar radiation.
It can be estimated from the energy emitted by and the position of the sun.
The sunlight reaching a horizontal plane on earth per unit area at time $t$, is called global horizontal irradiance $I_t\in\R_{0}^+$.
It includes direct normal irradiance, which originates directly from the sun, and diffuse irradiance, which includes scattered and ground reflected components, \cite{ISO9488}.

\begin{figure}[tbp]
	\centering
	\includegraphics{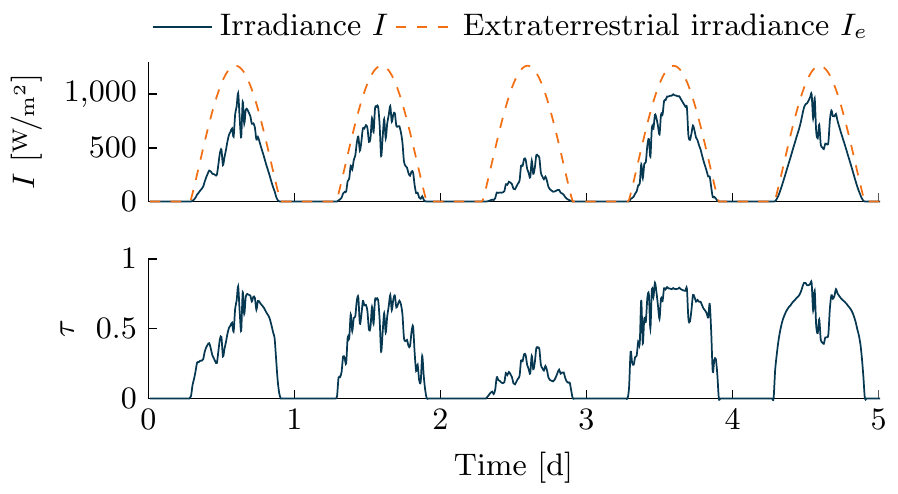}%
	\caption[Irradiance, Transmissivity]{Global horizontal irradiance, extraterrestrial irradiance and transmissivity over a duration of \unit[5]{d}.
	The irradiance was taken from \cite{ARM}, the extraterrestrial irradiance was estimated using \cite{RA2004}.
	 \label{fig:solarIrradiance}}
\end{figure}

For forecasting, it can be beneficial to normalize the solar irradiance $I_t$ by the extraterrestrial irradiance $I_{\text{e},t}\in\R_{0}^+$ (see Fig.~\ref{fig:solarIrradiance}).
This yields to transmissivity defined as
$
	\tau_t = \nicefrac{{I_t}}{{I}_{\text{e},t}}
$
\cite{QiaoZeng2012}.
In this work, the solar position algorithm from \cite{RA2004} is used to estimate the extraterrestrial irradiance $I_\text{e}(t)$.
It only requires the position of the respective surface to calculate the zenith angle $\zeta_t\in[0, 2\pi)$, i.e., the incidence angle of the sun light on a horizontal plane on the earth's surface \cite{Quaschning2009}.
Using $\zeta_t$, the extraterrestrial irradiance can be determined by
$
	I_{\text{e},t} = \epsilon_t I_{\text{s}} \cos(\zeta_t),
$
where $\epsilon_t$ is a correction factor and $I_{\text{s}} = \unitfrac[1360.8]{W}{m^2}$ is the solar constant.


\section{Time-Series Forecasting}
\label{sec:timeSeriesForecasting}

A time series is a collection of \inNLargerZero{n} chronologically ordered observations $y_1, y_2, \ldots, y_n$.
In this collection, every element $y_t$ refers to an observation performed at a time instant $t = 1, \ldots, n$.
In this work, we focus on univariate forecasts, i.e., forecasts using only historic irradiance to forecast future irradiance.
Thus, all elements $y_t\in\mathbb{R}$ for $t = 1, \ldots, n$ are real-valued scalars.

Broadly speaking, a forecasting method is a procedure to estimate future values $\hat{y}_{t+j|t}$ of a time series based on past values $Y_t \in \mathbb{R}^n$, i.e.,
$
	\hat{y}_{t+j|t} = f(Y_t).
$
Here, $\hat{y}_{t+j|t}$ refers to the value at prediction step $\inNLargerZero{j}$ with last known value at time $t\in\mathbb{N}$.
In case of univariate forecasting, $Y_t$ only includes the present and past values of the time series being forecast, i.e.,
$
	Y_t = [y_{t} ~ y_{t-1} ~ \cdots ~ y_{t-n+1}]\tran
$.

The vector $Y_t$ can also be based on non-consecutive lags, e.g., to represent seasonal behavior.
An example for such a vector is
$
	Y_t = [y_{t} ~ y_{t-s+1} ~ y_{t-s}]\tran,
$
where \inNLargerZero{s} is the length of the season, e.g., one day.
In this example,
the present observation, $y_t$,
the observation from the last day, $y_{t-s+1}$, at the same time as the predicted value, $\hat{y}_{t+1}$,
and from the current time minus \unit[24]{h}, $y_{t-s}$, are included.

\subsection{Multi-Step Forecasts}
\label{sec:multipleStepForecasts}
\begin{subequations}
To create multi-step ahead forecasts, a recursive strategy can be applied \cite[Section 4.]{Bontempi2013}.
The forecasting model is repeatedly used for one step forecasts, where the elements of $Y_t$ are adapted at each step until prediction step \inNLargerZero{J} is reached.
For example, if $Y_t$ includes the last three data points, $Y_t = [y_{t} ~ y_{t-1} ~ y_{t-2}]^{T}$
then the procedure for $J = 12$ prediction steps would be
\begin{align}
	\hat{y}_{t+1|t}		& = f(Y_t) = f([y_{t} \quad y_{t-1} \quad y_{t-2}]\tran), \\
	\hat{y}_{t+2|t}		& = f(\widehat{Y}_{t+1|t}) = f([\hat{y}_{t+1|t} \quad y_{t} \quad y_{t-1}]\tran), \\
						& \mathrel{\makebox[\widthof{=}]{\vdots}} \nonumber\\
	\hat{y}_{t+12|t}	& = f(\widehat{Y}_{t+11|t}) = f([\hat{y}_{t+11|t} \quad \hat{y}_{t+10|t} \quad \hat{y}_{t+9|t}]\tran).
\end{align}
For seasonal models, the vector  $\widehat{Y}_{t+j|t} \in \mathbb{R}^n$ is equally adapted at each step in a similar manner.
\end{subequations}

\subsection{Evaluation of Forecast Accuracy}
\label{sec:evaluation}

\subsubsection{Handling of Night Data.}
\label{sec:handlingOfNightData}

Forecasts of data points during the night, i.e., of zero irradiance values, were not included in the evaluation of the forecast accuracy.
Therefore, observations with a zenith angle
$
	\zeta_t \leq 90.83^\circ
$
were excluded based on \cite{Meeus1991}.
The angle $\zeta_t$ was taken from the data sets if available.
Otherwise it was estimated using \cite{RA2004}.

\subsubsection{Root Mean Square Error (RMSE).}\acused{rmse}
\label{sec:rmse}
For each prediction step $j = 1,\ldots, J$ over prediction horizon $J$, the \ac{rmse} is calculated to analyze the error of each step separately.
Assuming that \inNLargerZero{m} forecasts were performed, the \ac{rmse} of prediction step $j$ is
\begin{equation}
	\label{eq:rmse}
	\operatorname{RMSE}_j = \textstyle\sqrt{\frac{1}{m} \sum_{t=1}^{m} \left(\hat{y}_{t+j|t}-y_{t+j}\right)^2}.
\end{equation}


\section{Forecasting Methods}
\label{sec:forecastingMethods}
This section presents the forecasting methods employed in this work.
First, the persistence model used as reference is discussed.
Then, \ac{arima} models and \acl{nnr} are illustrated.

\subsection{Persistence Models}
\label{sec:persistenceModels}
The basic idea of the persistence model is that future values are assumed to be equal to known past values.
In this work, the simple model, $\hat{y}_{t+1|t} = y_t$ is used.
It is also possible to consider a seasonal persistence model using, e.g., data from the previous day, week or year.
This can be described by $\hat{y}_{t+j|t} = y_{t+j-s}$ where $s$ is the seasonal period.

\subsection{Autoregressive Integrated Moving Average (ARIMA) Models}\acused{arima}
\ac{arima} models are widely used for time-series forecasting.
Future values are estimated using a linear combination of previously observed values and forecasting errors.
Also, differencing can be applied to obtain stationary data.
An \ac{arima} model can be described by \cite{BoxJenkinsReinsel1994}
\begin{equation}
	\Phi(B) \nabla^d y_{k} = \theta_0 + \Theta(B) e_{t},
\end{equation}
where
$B$ is the backwards shift operator with $B^m y_{t} = y_{t-m}$,
and
$\nabla$ is the backwards difference with $\nabla^{d} y_{t} = (y_{t}-y_{t-1})^{d}$, \inNLargerEqualZero{d}.
Furthermore,
$\Phi(B) = 1 - \phi_1 B^{1} - \ldots - \phi_p B^{p}$
with $\phi_1, \ldots, \phi_p \in \mathbb{R}$, \inNLargerEqualZero{p}
is the autoregressive part,
and
$\Theta(B) = 1 - \theta_1 B^{1} - \ldots - \theta_q B^{q}$
with $\theta_1, \ldots, \theta_q \in \mathbb{R}$, \inNLargerEqualZero{q}
the moving average part.
Moreover, $e_{t}$ is the difference between measured value and forecast.

To fit a model to data that exhibits seasonality, \ac{arima} models can be extended by seasonality of period $s$.
These, so called \ac{sarima} models, can be described by
\begin{equation}
	\Phi(B) \mathbf{\Phi}(B) \nabla^d \nabla^D_s y_{t} = \theta_0 + \Theta(B) \mathbf{\Theta}(B) e_{t}, \label{eq:sarima}
\end{equation}
where
$\mathbf{\Phi}(B) = 1 - \Phi_1 B^{s} - \ldots- \Phi_P B^{P\,s}$
with $\Phi_1, \ldots, \Phi_P \in \mathbb{R}$, \inNLargerEqualZero{P}
is the seasonal autoregressive part
and
$\mathbf{\Theta}(B) = 1 - \Theta_1 B^{s} - \ldots - \Theta_Q B^{Q\,s}$
with $\Theta_1, \ldots, \Theta_Q \in \mathbb{R}$, \inNLargerEqualZero{Q}
the seasonal moving average part.
Moreover, $\nabla^D_{s} y_{t}= \left(y_{t} -y_{t-s}\right)^D$, \inNLargerEqualZero{D} represents seasonal differencing.

For the implementation of the \ac{arima} models, the econometrics toolbox in MATLAB~2015b was used.
Within the toolbox, maximum likelihood estimation is used to find the model parameters.

\subsection{Nearest Neighbor Regression (NNR)}\acused{nnr}
\label{sec:nnr}

In what follows, based on \cite{Altman1992} we introduce \ac{nnr}.
In \ac{nnr} the pattern of historic data at time instant $t$, $Y_t\in \mathbb{R}^n$, is compared to previous patterns $Y_i \in \mathbb{R}^n$, $i = 0, \ldots, N-1$ in the reference sample 
$\mathbb{D} = \left\{ (Y_0, y_1), (Y_1, y_2), \ldots, (Y_{N-1}, y_N) \right\}$,
$N\in\mathbb{N}$
to forecast $\hat{y}_t$.

Although \ac{nnr} allows to use arbitrary elements in $Y_i$, e.g., $Y_i = [y_{i-13} ~ y_{i-111} ~ y_{i-2678}]\tran$, we follow the notion of autoregressive and seasonal autoregressive lags of (seasonal) \ac{arima} models to enable a comparison of both methods.
For \ac{nnr} this means that according to the number autoregressive lags $p$ the vectors $Y_i$ in $\mathbb{D}$ are formed.
Thus, analog to \ac{arima} models the entries of the \ac{nnr} reference sample with $p = 3$ autoregressive lags has the form
$
	Y_{i} = \left[	y_{i} ~ y_{i-1} ~ y_{i-2}\right]\tran.
$

The same holds for the seasonal autoregressive lags.
In a similar fashion as in \eqref{eq:sarima}, the seasonal autoregressive part is multiplied with the autoregressive part.
Consequently, a reference sample with $p = 3$ autoregressive lags, $P = 2$ seasonal autoregressive lags and a season of $s=10$ results in
\[
	Y_{i} = \left[	y_{i} \quad 	y_{i-1} \quad 	y_{i-2} \mid
					y_{i-9} \quad 	y_{i-10} \quad 	y_{i-11} \quad y_{i-12} \mid
					y_{i-19} \quad 	y_{i-20} \quad 	y_{i-21} \quad y_{i-22}\right]\tran.
\]
Thus, as in \eqref{eq:sarima}, the elements of the previous seasons that are associated with $\hat{y}_{i+1|i}$
(here $y_{i-9}$ and $y_{i-19}$) are included in $Y_i$.
This is due to the zero order term, i.e., the $1$, in $\Phi(B)$ and $\mathbf{\Phi}(B)$.

A simple forecasting model can be obtained by combining the \inNLargerZero{k} elements in $\mathbb{D}$ with the smallest distance to $Y_t$.
Using the set of $k$ nearest neighbors of $Y_t$, $\mathbb{K}(Y_t) \subseteq \mathbb{D}$, this can be written as
\begin{equation}
	\hat{y}_{t+1|t} = f(Y_t, \mathbb{D}) = \nicefrac{1}{k}\textstyle\sum_{Y_{i} \in \mathbb{K}(Y_t)} y_{i+1}. \label{eq:nnRegressionFunction}
\end{equation}
Note that the number of neighbors $k$ can be fixed or determined by a maximum distance $\varepsilon \in \mathbb{R}^+$.
With distance $d(Y_i, Y_t)$, the set of neighbors closer than $\varepsilon$ to $Y_t$ is $\mathbb{K}(Y_t) = \{Y_i \in \mathbb{D} \mid d(Y_i, Y_t) \leq \varepsilon \}$.
Note that in this work, the Euclidean norm is used as distance, i.e., $d(Y_i, Y_t) = \|Y_{i} - Y_{t}\|_2$.

The model \eqref{eq:nnRegressionFunction}, can be modified using a weighted average.
In this work, weights inverse to the distance between $x_t$ and the neighbor $x_{i}$, have been considered, i.e.,
\begin{equation}
	\hat{y}_{t+1|t} = f(Y_t, \mathbb{D}) =
	\begin{cases}
		\nicefrac{1}{k}\sum_{Y_{i} \in \mathbb{K}(Y_t)}\frac{1}{d(Y_i, Y_t)} y_{i+1}, & \text{if } d(Y_i, Y_t)\neq 0 ~ \forall Y_i \in \mathbb{K}(Y_t), \\
		y_{l+1}, & \text{if } \exists Y_l \in \mathbb{K}(Y_t) \text{ with } d(Y_t, Y_l)= 0.\\
	\end{cases}
	\label{eq:weightDistance}
\end{equation}

Although, \eqref{eq:nnRegressionFunction} requires no training, the model and the selected reference sample will be referred to as trained model for simplicity.
For implementation, the Statistical Learning Toolbox for MATLAB~\cite{Lin2006} was used.
Note that in this work, approximated nearest neighbors which is faster than the exact nearest neighbors algorithm is used as no significant difference in the forecast accuracy between the two could be observed.


\section{Hyperparameter Search}
\label{sec:modelSelection}

\begin{figure}[t]
	\centering
	\includegraphics{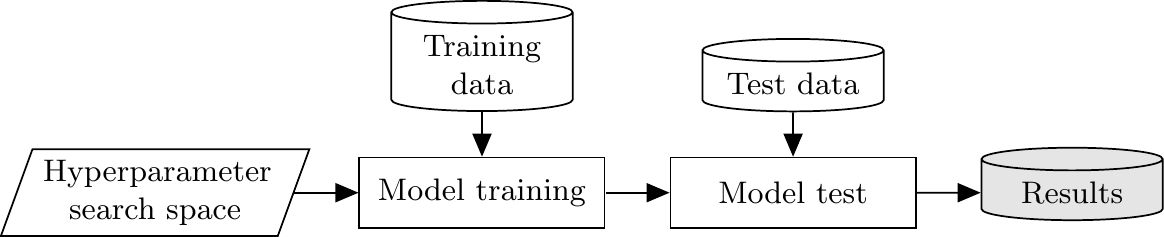}%
	\caption{Scheme used to obtain forecast accuracy data for different models.}
	\label{fig:flowChart}
\end{figure}

We aim to compare the forecast accuracy for different forecasting model structures and different sets of historic data.
Therefore, an exhaustive search was performed to evaluate what models approximate the test data best.
In our search, different model structures are trained using different sets of historic data.
To identify model structure and historic data, hyperparameters are used.
Moving along Fig.~\ref{fig:flowChart}, in this section first training and test data are discussed.
Then, the hyperparameter search space is posed.
Finally, training and test of the models is illustrated.

\subsection{Training and Test Data}
\label{sec:data}

\begin{figure}[h]
	\includegraphics{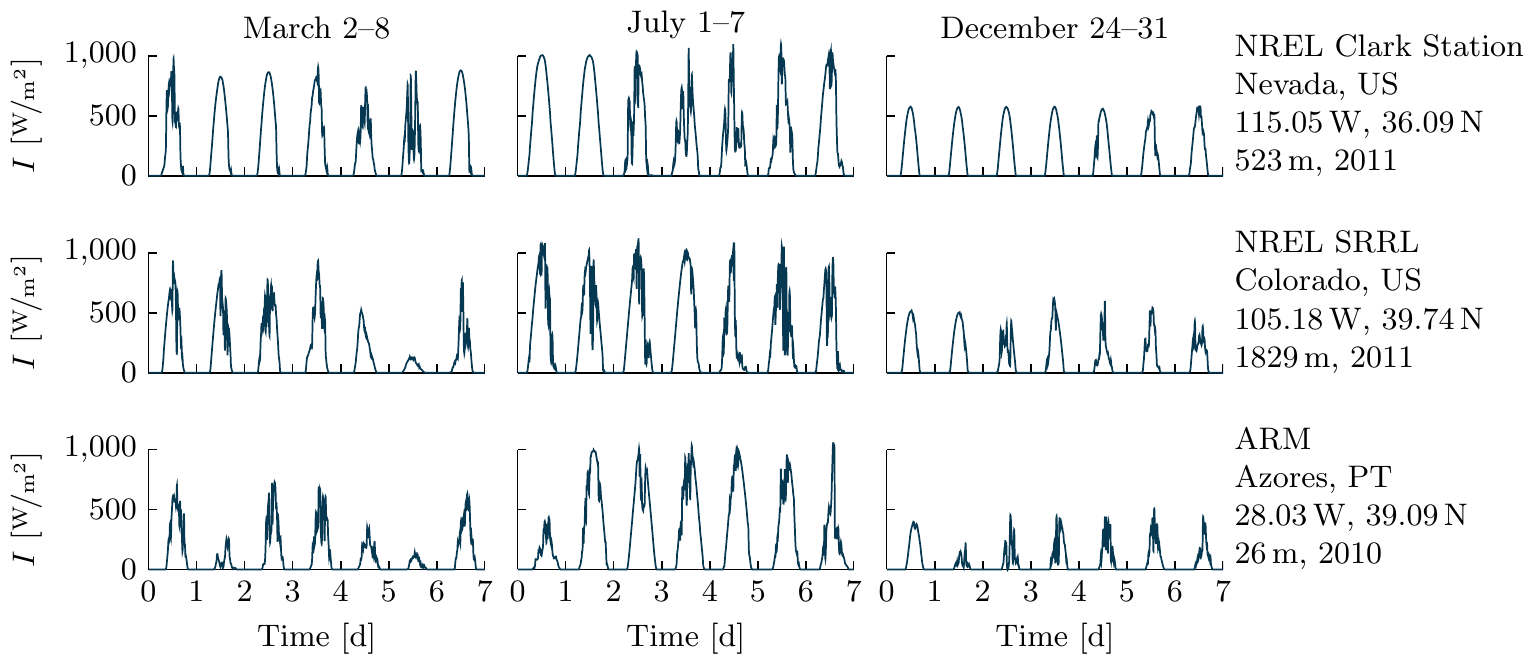}%
	\caption[Test Data]{Measured solar irradiance data used to test the trained models.}
	\label{fig:testdata}
\end{figure}

For training and test data sets \cite{NRELNevada,NRELSRRL,ARM} were used.
From each data set three different seasons were selected to include different climatic situations in the hyperparameter search.
For every location and every season, i.e., for 9 data sets, the observations were divided into training data (up to 2~months) and independent test data (1~week) that was not used for training.
The test data was chosen to always follow directly the data used for training.
As shown in Fig.~\ref{fig:testdata}, the nine weeks of test data include different climatic situations, e.g.,
a sunny week in December from the \ac{nrel} Clark Station,
a cloudy week in March from the \ac{arm} Facility, and
various weeks with sunny, foggy and cloudy days.

Note that all data sets \cite{NRELNevada,NRELSRRL,ARM} have a temporal resolution of \unit[1]{min}.
The \unit[15]{min} average values $y_1, \ldots, y_N$ were derived from the original time series $y'_1, \ldots, y'_{15N}$ as $y_t = \nicefrac{1}{15} \textstyle\sum_{q=1}^{15} y'_{q+15(t-1)}$.

\subsection{Hyperparameters}
\label{sec:hyperparameters}

\subsubsection{Data Hyperparameters.}
\label{sec:dataHyperparameters}

The hyperparameters that concern the training data are independent of the forecasting method and the model structure.
We considered modifications of training data by
\begin{enumerate*}
	\item pre- and postprocessing,
	\item handling of night data ans
	\item number of data points used.
\end{enumerate*}

1. Regarding the pre- and postprocessing, two approaches were considered.
(a) The time series of irradiance, $I_t$, is directly used to train the models without further preprocessing.
(b) Transmissivity $\tau_t = \nicefrac{I_t}{I_{\text{e},t}}$ is derived from $I_t$ and extraterrestrial irradiance~$I_{\text{e},t}$.
Then, the models are trained to forecast $\hat{\tau}_{t+j|t}$.
Using $\hat{\tau}_{t+j|t}$ the irradiance forecast is then calculated via $\hat{I}_{t+j|t} = I_{\text{e},t} \, \hat{\tau}_{t+j|t}$.

2. Three cases related to the handling of data points at night were considered:
(a) include all data points from day and night,
(b) only include data points between \unit[5]{am} and \unit[8]{pm} local time, and
(c) only include data points between sunrise and sunset, i.e., observations with $\zeta_t > 90.83^\circ$ (see Section~\ref{sec:handlingOfNightData}).
As it can be cumbersome to implement models with seasonality for a varying number of data points per day, case (3) was not considered for seasonal models.

3. Regarding the number of training data points, $1$, $3$, $7$, $14$, $30$, and $60$ days of data were considered.
For seasonal models, the data was chosen to include at least $14$ days.

\subsubsection{Model Hyperparameters.}


\begin{table}[tb]
\caption{Sets of hyperparameters for different forecasting methods considered in search.}
\label{tab:parameter}
\begin{subtable}[t]{0.49\textwidth}
	\centering
	\caption{ARIMA models.}
	\label{tab:parameterArima}
	\begin{tabular}{p{5cm}p{2.2cm}}
		\toprule
		Autoregressive lags $p$ & $0, 1, \ldots, 10$ \\
		Moving average lags $q$ & $0, 1, \ldots, 10$ \\
		Differencing $d$		& $0, 1, 2$ \\
		\bottomrule
	\end{tabular}
\end{subtable}
%
\hfill
\begin{subtable}[t]{0.49\textwidth}
\centering
	\caption{NNR.}
	\label{tab:parameterNnr}
	\begin{tabular}{p{4.4cm}p{2.8cm}}
		\toprule
		Autoregressive lags $p$		& $1, 2, \ldots, 20$ \\
		Weight 						& uniform, \newline inverse to distance \\
		Number of neighbors $k$ 	& $1, 2, \ldots, 20$ \\
		Threshold\footnotemark[3] $\varepsilon$ & $0.01, 0.05, 0.1, 0.5, 1$ \\
		\bottomrule
	\end{tabular}%
\end{subtable}

\hspace{\fill}

\begin{subtable}[b]{0.49\textwidth}
	\centering
	\caption{SARIMA models.}
	\label{tab:parameterSarima}
	\begin{tabular}{p{5cm}p{2.2cm}}
		\toprule
		Autoregressive lags	$p$				& $0, 1, 3$ \\
		Moving average lags $q$				& $0, 1, 3$ \\
		Seasonal autoregressive lags $P$ 	& $0, 1, 2, 3$ \\
		Seasonal moving average lags $Q$	& $0, 1, 2, 3$ \\
		Differencing $d$ 					& $0, 1$ \\
		Seasonal differencing $D$			& $0, 1$ \\
		\bottomrule
	\end{tabular}
\end{subtable}
\hfill
\begin{subtable}[b]{0.49\textwidth}
\centering
	\caption{SNNR.}
	\label{tab:parameterSeasonalNnr}
	\begin{tabular}{p{4.4cm}p{2.8cm}}
		\toprule
		Autoregressive lags $p$				& $1, 2, \ldots, 11$ \\
		Seasonal autoregressive lags $P$	& $1, 2, \ldots, 7$ \\
		Weight 							& uniform, \newline inverse to distance \\
		Number of neighbors $k$ 		& $1, 2, \ldots, 20$ \\
		Threshold\footnotemark[3]  $\varepsilon$ 	& $0.01, 0.05, 0.1, 0.5, 1$ \\
		\bottomrule
	\end{tabular}%
\end{subtable}
\end{table}

\footnotetext[3]{Multiplied by $\unitfrac[1360.8]{W}{m^2}$ for irradiance.}

For every forecasting method, the model structure can be uniquely described using model specific hyperparameters.
Note the difference to model parameters of a trained model.
For example, the number of autoregressive lags $p$ is a hyperparameter.
The coefficients of the autoregressive part, $\phi_1, \ldots, \phi_p$, of a trained \ac{arima} model are model parameters.
The subsets of hyperparameters considered in the search for are shown in Table~\ref{tab:parameter}.
Note that in Tables~\ref{tab:parameter}(\subref{tab:parameterNnr}) and~\ref{tab:parameter}(\subref{tab:parameterSeasonalNnr}) either the number of neighbors $k$ or the maximal distance to the neighbors $\varepsilon$ is used.
Further note that the season $s$ was chosen to be \unit[1]{d} for \ac{sarima} and \ac{snnr}.

\subsection{Model Training}
For each combination of hyperparameters described in Section~\ref{sec:hyperparameters}, we trained nine models, i.e., one for each location and each season (see Section~\ref{sec:data}).
This lead to more than 250,000 trained forecasting models with different hyperparameters.
Note that some combinations of hyperparameters did not result in suitable models.
Especially for \ac{arima}, no stable model could be derived for certain hyperparameters.
Furthermore, for \ac{sarima} \unit[3.3]{\%} of the hyperparameters were excluded, mostly due to very long training procedures that rendered them unsuitable for practical use.

\subsection{Model Test}
\label{sec:modelTest}

In the model test, the performance of the trained models was evaluated.
Therefore, every trained model was tested by forecasting irradiance of the unknown test data (see Fig.~\ref{fig:testdata}) that followed the training data.
Due to zero irradiance, data points during the night were excluded from the evaluation based on the zenith angle $\zeta_t$ (see Section~\ref{sec:handlingOfNightData}).
For each of the data point in the test data, a $12$ step ahead, i.e., \unit[3]{h}, prediction was obtained.
Using the forecast values and the test data, the $\operatorname{RMSE}_j$ (see Section~\ref{sec:rmse}) was then calculated for $j=1, \ldots, 12$ and stored.
In the next section, we will analyze the resulting $\operatorname{RMSE}_j$ for every combination of hyperparameters, location and season to identify models that are likely to provide good forecasts.


\section{Analysis}
\label{sec:analysis}
In this section, we analyze the results of the hyperparameter search in Section~\ref{sec:modelSelection}.
Our goal is to draw conclusions that help to reduce the space of future hyperparameter searches.
More precisely, we aim to find forecasting methods and ranges of hyperparameters that increase the probability of finding a suitable model for unseen data.
Before starting the analysis, some remarks are posed.

\begin{remark}\label{rem:referenceModel}
	As reference, we use the persistence model on transmissivity as it outperformed the persistence model on irradiance for all test data sets.
	In the reference model historic transmissivity is derived using $\tau_t = \nicefrac{I_t}{I_{\text{e},t}}$.
	Then a persistence forecast is performed, i.e., we set $\hat{\tau}_{t+j|t} = \tau_t$ for $j = 1, \ldots, J$.
	Finally, the irradiance forecast is determined as $\hat{I}_{t+j|t} = \hat{\tau}_{t+j|t} \, I_{\text{e},t+j}$ for $j = 1, \ldots, J$.
\end{remark}

\begin{remark}
	The style of the box plots used throughout this work is illustrated in in Fig.~\ref{fig:boxPlotExplaination}.
	Here, $m$ marks the median.
	The box around the median contains all data from the 25th ($q_1$) to the 75th ($q_3$) percentile.
	The left whisker marks the lowest occurring value within $q_1 - 1.5 (q_3 - q_1)$ and the right whisker marks the highest occurring value within $q_3 + 1.5 (q_3 - q_1)$.
	Due to numerous outliers, only every 100th outlier is shown.
	To assure that the forecast with the highest accuracy is considered the lowest outlier is always included.

	\begin{figure}[h]
		\centering
		\includegraphics{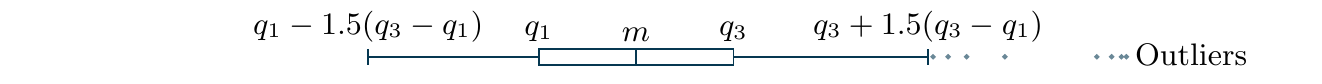}%
		\caption[/boxPlotExplaination.tex]{Box plot with median $m$, quartiles $q_1$ and $q_3$, whiskers and outliers.}
		\label{fig:boxPlotExplaination}
	\end{figure}
\end{remark}

\begin{remark}
	Most distributions of the forecast accuracy discussed in this section are similar for the different test data sets shown in Fig.~\ref{fig:testdata}.
	Therefore, most distributions were combined in one plot that includes the values of all locations.
\end{remark}

\begin{remark}
	Throughout this paper, the forecast at prediction steps $j = 1, 4, 12$ are analyzed to approximate the evolution of the forecast accuracy over prediction horizon $J=12$.
	For the \unit[15]{min} sampling time considered, they correspond to a forecast \unit[15]{min}, \unit[1]{h} and \unit[3]{h} ahead.
\end{remark}

\subsection{Forecasting Method}
\label{sec:forecastMethod}

\begin{figure}[htbp]
	\centering
	\includegraphics{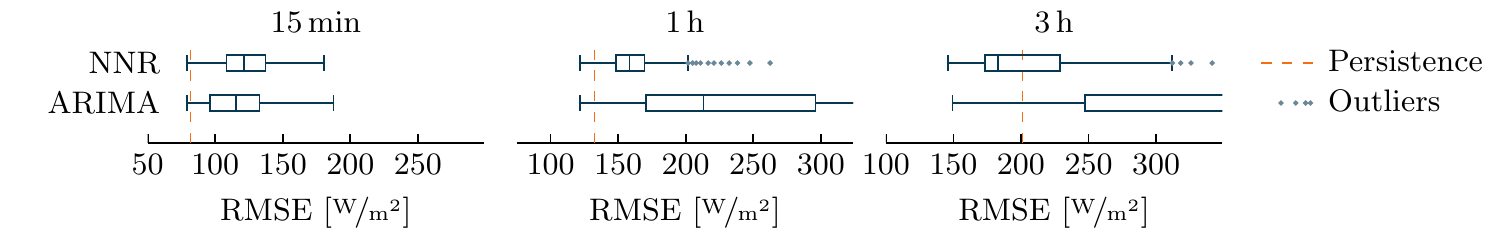}%
	\caption[/boxPlotArimaNnr.tex]{\ac{rmse} for \unit[15]{min}, \unit[1]{h} and \unit[3]{h} ahead predictions with \ac{arima} and \ac{nnr} models.}
	\label{fig:boxPlotArimaNnr}
\end{figure}

We first compare the forecasting methods \ac{arima} and \ac{nnr} along with the persistence model on transmissivity that acts as a reference (see Remark~\ref{rem:referenceModel}).
In Fig.~\ref{fig:boxPlotArimaNnr}, box plots illustrating the distributions of forecast accuracies of all models included in the search are shown.

The box plots in Fig.~\ref{fig:boxPlotArimaNnr} show that for predictions \unit[15]{min} ahead, almost no model outperforms the persistence model.
For a prediction step of \unit[1]{h} some models outperform the persistence forecaster.
Furthermore, the \ac{rmse} of forecasts performed with \ac{nnr} varies less than the \ac{rmse} of \ac{arima} forecasts.
This indicates a higher probability of finding a good forecasting model using \ac{nnr}.
For predictions of \unit[3]{h} ahead, the potential improvement over the persistence model is significant for all methods.
As already observed for \unit[1]{h} ahead predictions, the \ac{rmse} of \ac{nnr} varies less.
Additionally, the 25th percentile is remarkably lower for \ac{nnr}.
The highest forecast accuracy achieved is very alike for \ac{nnr} and \ac{arima} for all prediction steps.
Still, \ac{nnr} achieves much lower medians and lower 25th and 75th percentiles.
Consequently, the probability of finding a suitable \ac{nnr} model is much higher.
Therefore, \ac{arima} and \ac{sarima} models will be disregarded in the next steps of the analysis.
Consequently, in what follows we only investigate how the search spaces for \ac{nnr} and \ac{snnr} can be reduced to increase the probability of finding an accurate forecasting model.

\subsection[Preprocessing]{Preprocessing of \ac{nnr} and \ac{snnr}}
\label{sec:preprocessing}

\begin{figure}[ht]
	\centering
	\includegraphics{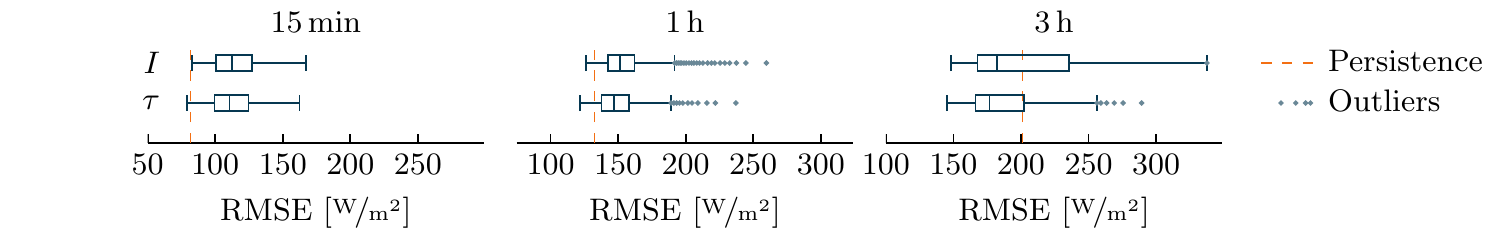}%
	\caption[Box plot nnr preprocessing]{%
	\ac{rmse} of \ac{nnr} and \ac{snnr} using different preprocessing for all locations and seasons.}
	\label{fig:boxPlotGhiCsi}
\end{figure}

\begin{figure}[h!]
	\centering
	\includegraphics{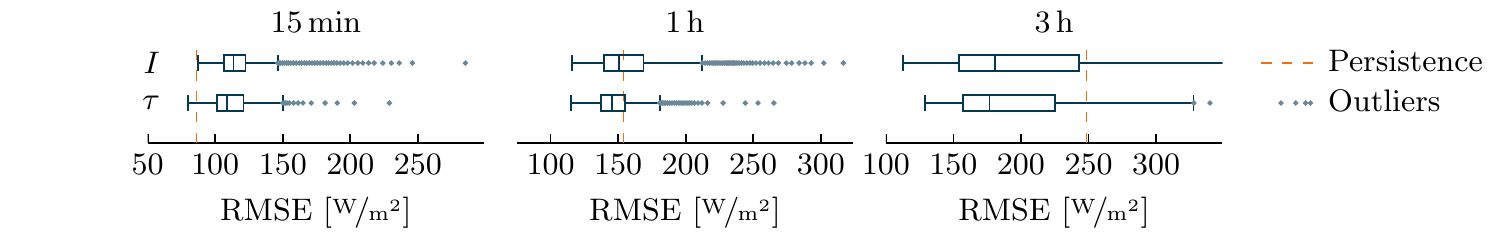}%
	\caption[Box plot nnr preprocessing exception]{\ac{rmse} of \ac{nnr} and \ac{snnr} using different preprocessing at \ac{nrel} Clark Station, March 2--8.}
	\label{fig:boxPlotNnrPreprocessingException}
\end{figure}

As stated in Section~\ref{sec:dataHyperparameters}, models using irradiance or the normalized transmissivity values were considered in the search.
In Fig.~\ref{fig:boxPlotGhiCsi} the forecast accuracy of both approaches is shown.
It can be observed that the box plots for \unit[15]{min} forecasts are very alike.
Also, little difference can be seen for \unit[1]{h} ahead predictions.
However, for \unit[3]{h} ahead predictions the transmissivity based forecasts often outperform the irradiance based forecasts.
The only exception from this observation could be found for data from the \ac{nrel} Clark Station during March 2--8 (see Fig.~\ref{fig:boxPlotNnrPreprocessingException}), where the best models use irradiance data.
Still the lower whiskers and the medians in this case are very alike.

The box plots indicate that using transmissivity yields a higher probability of finding an accurate model.
Although for data set \cite{NRELNevada} a better irradiance model for the \unit[3]{h} prediction could be found, the overall distribution in this case shows no significant additional difference to Fig.~\ref{fig:boxPlotGhiCsi}.
As the single best model is hard to find, we are confident that using transmissivity yields a higher probability of finding a good model.
Therefore, irradiance forecasts are excluded from the reduced search space.

\subsection[Handling of Night Data]{Handling of Night Data of \ac{nnr} and \ac{snnr}}
\label{sec:nightData}

As stated in Section~\ref{sec:modelTest}, data points during the night were not included in the evaluation of the forecast accuracy.
However, in some cases they were used in the data of the reference sample to allow for smooth transitions between two days.
As stated in Section~\ref{sec:data}, we considered training data that
(a) includes the data of the entire day (and night),
(b) only includes data points from sunrise to sunset and
(c) only includes data points from \unit[5]{am} to \unit[8]{pm}.

\begin{figure}[t]
	\centering
	\includegraphics{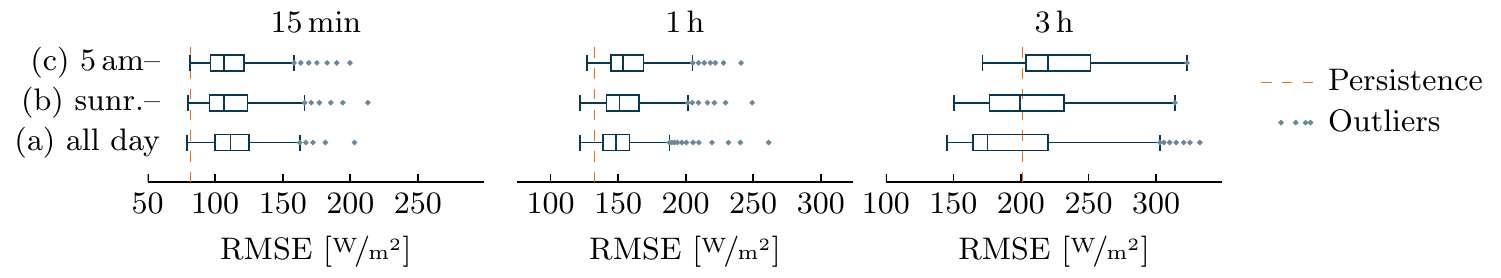}%
    \caption[Box Plot NNR Night Data]{%
	\ac{rmse} of \ac{nnr} and \ac{snnr} for different handling of night data: (a)~entire day, (b)~sunrise to sunset and (c)~\unit[5]{am} to \unit[8]{pm}.}
	\label{fig:boxPlotNightData}
\end{figure}

In Fig.~\ref{fig:boxPlotNightData}, the forecast accuracy for different handling of night data is shown.
It can be observed that the forecast accuracy is similar for the \unit[15]{min} forecast.
For forecasts \unit[1]{h} ahead, the difference between the approaches remains small.
However, for forecasts \unit[3]{h} ahead, using all data points for training, i.e., including night values, outperforms the other approaches.
Therefore, all data points including night data are considered in the reduced search space.

\subsection[Size of the Reference Sample]{Size of the Reference Sample of \ac{nnr} and \ac{snnr}}
\label{sec:sizeSample}

The reference sample is chosen such that it includes that last values of historic data up to the most recent data point.
A large reference sample, i.e., a set $\mathbb{D}$ with a high number of elements, therefore includes data that reaches further into that past than a small reference sample.

\begin{figure}[ht]
	\centering
	\includegraphics{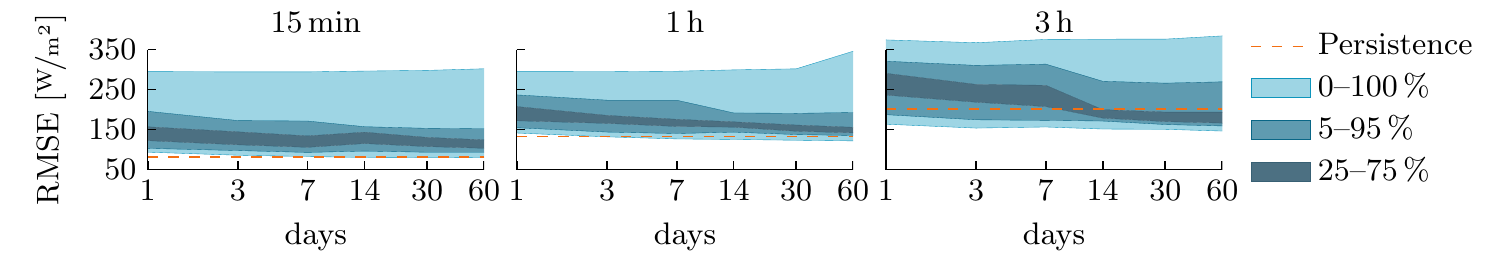}%
	\caption[Fan chart NNR size data training]{%
	\ac{rmse} of \ac{nnr} and \ac{snnr} using different sizes of the reference sample.}
	\label{fig:sizeDataTraining}
\end{figure}

In Fig.~\ref{fig:sizeDataTraining}, the forecast accuracy using a reference sample of \unit[1]{d}, \unit[3]{d}, and \unit[7]{d} as well as two weeks (\unit[14]{d}), three weeks (\unit[21]{d}), one month (\unit[30]{d}) and two months (\unit[60]{d}) is shown.
The plots for all prediction steps shows a small decrease of the 25th and the 75th percentile as well as the lowest \ac{rmse} for an increasing number of elements in the reference sample.
Considering the largest \ac{rmse} of each prediction step, the inverse effect can be observed.
In general little difference can be observed in the forecast accuracy, when varying the size of the reference sample.
However, as the forecast accuracy increases using more data, a reference sample of \unit[60]{d} is used in the reduced search space.

\subsection[Autoregressive Lags]{Autoregressive (AR) Lags of \ac{nnr} and \ac{snnr}} \acused{ar}
\label{sec:analysisLags}
This section focuses on the \ac{ar} lags, i.e., the data points of the historic data used in the elements of the reference sample.
We differentiate between \ac{ar} and seasonal \ac{ar} lags.
The former refer to \unit[15]{min} steps prior to the first forecast time instant.
The latter refer to \unit[24]{h} steps, used to include the days prior to the forecast.
\begin{figure}[t]
	\centering
	\includegraphics{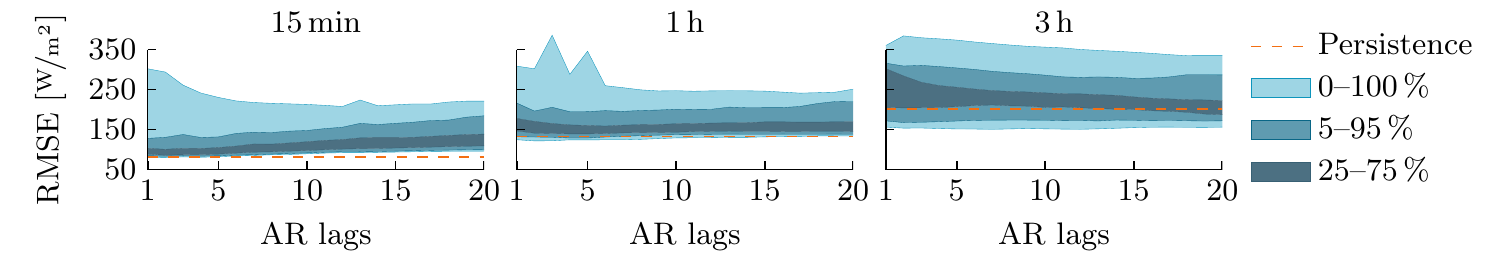}%
	\caption[fanChartNnrLags]{\ac{rmse} of \ac{nnr} using different \ac{ar} lags.}
	\label{fig:fanChartNnrLags}
\end{figure}
%
It can be observed in Fig.~\ref{fig:fanChartNnrLags} that for all prediction steps the models with the highest forecast accuracy has a small number of \ac{ar} lags.
However, no clear tendency can be identified that supports choosing a particular number of \ac{ar} lags.

\begin{figure}[ht]
	\centering
	\includegraphics{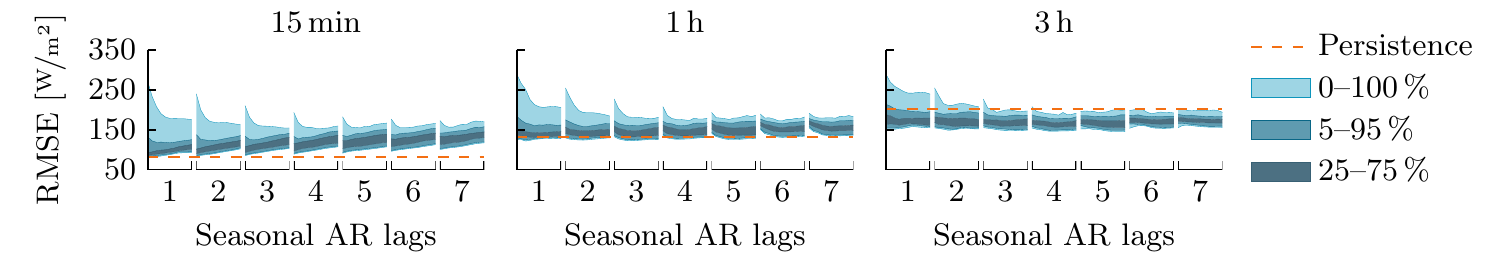}%
	\caption[fanChartNnrSeasonalLags]{%
	\ac{rmse} of \ac{snnr} using different seasonal \ac{ar} lags.
	For every seasonal \ac{ar} lag, the non-seasonal \ac{ar} lags increase from $1$ on the left side of the plot to $11$ on the right side of the small subplot.}
	\label{fig:fanChartNnrSeasonalLags}
\end{figure}

Fig.~\ref{fig:fanChartNnrSeasonalLags} shows the \ac{rmse} for different numbers of seasonal \ac{ar} lags.
Here, for every seasonal \ac{ar} lag, the number of \ac{ar} lags was increased from $1$ (left) to $11$ (right).
It can be observed that for \unit[15]{min} forecasts the \ac{rmse} slightly increases with the number of seasonal \ac{ar} lags.
Furthermore, for every seasonal \ac{ar} lag, the \ac{rmse} also increases with increasing number of non-seasonal \ac{ar} lags.
In contrast, for the \unit[3]{h} forecasts, the \ac{rmse} slightly decreases with the number of seasonal lags.
Comparing Figs.~\ref{fig:fanChartNnrLags} and~\ref{fig:fanChartNnrSeasonalLags}, shows that the variance decreases for the \unit[1]{h} and \unit[3]{h} forecasts with increasing number of seasonal lags.
This results in a higher probability of finding a good model for a higher number of seasonal \ac{ar} lags.
Unfortunately, the \unit[15]{min} forecasts do not show the same effect.
Still, as the \unit[15]{min} forecast barely outperform the persistence model it seems reasonable focus on the improvement for larger horizons and consider only seasonal models in the reduced search space.

\subsection[NNR Weights]{Weights of \ac{nnr} and \ac{snnr}}
\label{sec:weights}
As stated in Section~\ref{sec:hyperparameters} uniform weights and weights inverse to the distance to the neighbors were considered.
In the following, we analyze the effect of the distance to neighbors on the distribution of the \ac{rmse}.
\begin{figure}[bt]
	\centering
	\includegraphics{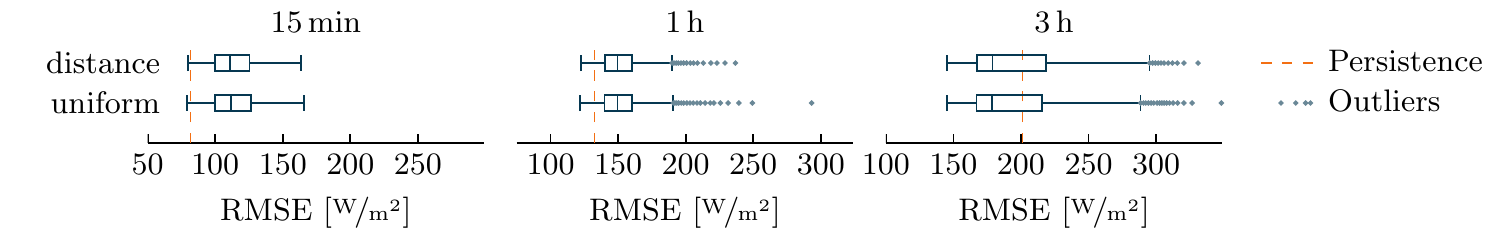}%
	\caption[Boxplot weights knn]{%
	\ac{rmse} using uniform weights and weights inverse to the distance of the neighbor.}
	\label{fig:boxPlotKnnWeights}
\end{figure}
%
As can be seen in Fig.~\ref{fig:boxPlotKnnWeights}, the box plots hardly differ for all prediction steps.
This indicates that there is no strong correlation between forecast accuracy and weights.
However, because of an easier implementation, uniform weights were chosen for the reduced search space.

\subsection[Definition of the Neighborhood]{Definition of the Neighborhood of \ac{nnr} and \ac{snnr}}
\label{sec:neighborhood}
As stated in Section~\ref{sec:nnr}, the neighborhood can be chosen to be a fixed number of neighbors, $k$, or a maximum distance, $\varepsilon$.
\begin{figure}[t]
	\centering
	\includegraphics{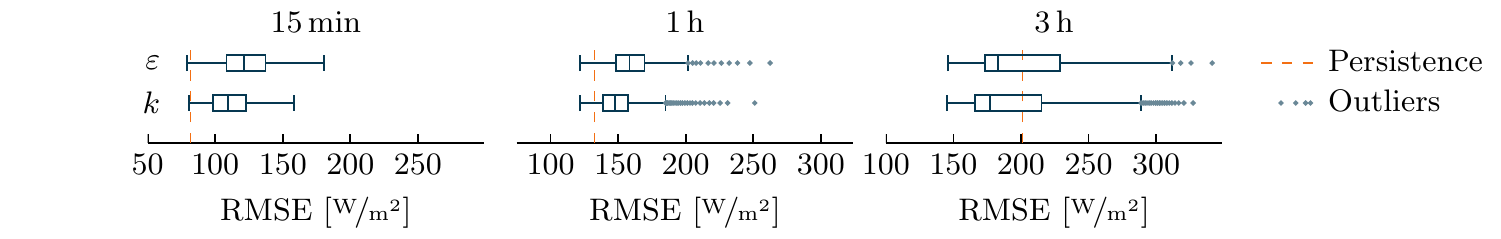}%
	\caption[BoxplotKnnKernel]{%
	\ac{rmse} using a threshold $\varepsilon$ or a fixed number of neighbors $k$ to define the neighborhood $\mathbb{K}$.}
	\label{fig:boxPlotKnnVsKernel}
\end{figure}
The comparison in Fig.~\ref{fig:boxPlotKnnVsKernel} shows that the highest accuracy achieved with both hyperparameters is very alike for all prediction steps.
However, the 25th and 75th percentile, the upper whisker and the median are lower for models using a fixed number of neighbors $k$.
Therefore, in the reduced search space, only models using a fixed number of neighbors $k$ are considered.

\subsection[Number of Neighbors]{Number of Neighbors $k$ of \ac{nnr} and \ac{snnr}}
\label{sec:numberNeighbors}

\begin{figure}[htb]
	\centering
	\includegraphics{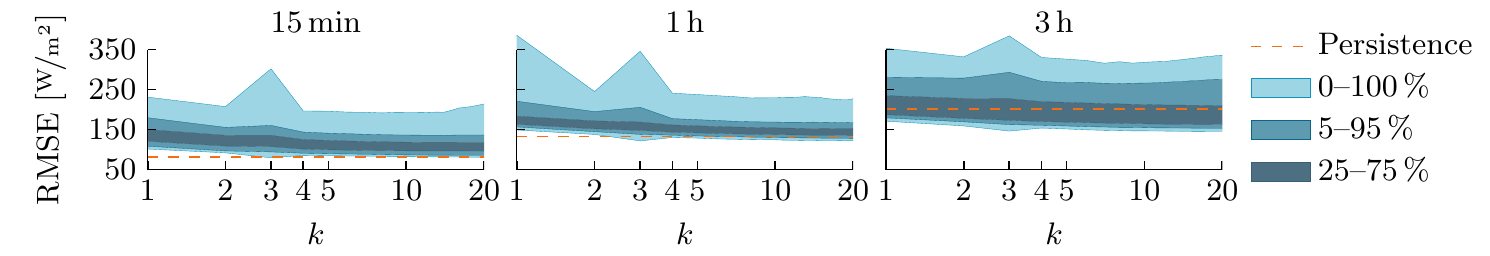}%
	\caption[Fan chart kNN - number of neighbours k.]{\ac{rmse} of \ac{nnr} and \ac{snnr} using a different number of neighbors $k$.}
	\label{fig:fanChartNumberOfNeighbors}
\end{figure}

In Fig.~\ref{fig:fanChartNumberOfNeighbors}, the forecast accuracy of all \acl{knn} models included in the search are shown.
It can be observed, that the forecast accuracy mostly changes using $1$ to $5$ neighbors.
The distribution of the \ac{rmse} becomes narrower for most cases with an increasing $k$.
Furthermore, the lowest \ac{rmse} decreases slightly for $k \geq 10$.
Therefore, models using between 10 and 20 neighbors are considered the reduced search space.

\subsection{Summary}
\label{sec:results:summary}

In summary, the following conclusions can be drawn from the analysis.
\begin{enumerate}
	\item The probability of finding a sufficiently accurate model to forecast irradiance is much higher for \ac{nnr} and \ac{snnr} than for \ac{arima} and \ac{sarima} models (see Section~\ref{sec:forecastMethod}).
	\item Data pre- and postprocessing of \ac{nnr} and \ac{snnr} models:
	\begin{enumerate}
		\item Use transmissivity instead of irradiance (Section~\ref{sec:preprocessing}).
		\item Include night data in the reference sample (Section~\ref{sec:nightData}).
		\item Use data from the past \unit[60]{d} in the reference sample (Section~\ref{sec:sizeSample}).
	\end{enumerate}
	\begin{samepage}
	\item Hyperparameters of \ac{nnr} and \ac{snnr} models: \label{enum:results:hyperparameters}
	\begin{enumerate}
		\item Seasonal models are beneficial (Section~\ref{sec:analysisLags}).
		\item Favor uniform weights over weights inverse to distance (Section~\ref{sec:weights}).
		\item Choose the neighbors using a fixed number $k$ instead of a maximum distance (Section~\ref{sec:neighborhood}).
		\item Search for models with 10 to 20 neighbors (Section~\ref{sec:numberNeighbors}).
	\end{enumerate}
	\end{samepage}
\end{enumerate}

Based on these findings the hyperparameter search space, i.e., the number of potential models could be reduced from more than 250,000 to less than 1,000.
The search space for suitable \ac{knn} forecasters now only includes the
number of autoregressive lags with range $1, 2, \ldots, 11$,
number of seasonal autoregressive lags with range $1, 2, \ldots, 7$ and
number of neighbors with range~$10, 11, \ldots, 20$.



\section{Conclusion}
A comparison of \ac{arima} and \ac{nnr} for short term solar forecasts showed that it is more likely to find good forecasting models using \ac{nnr}.
Based on this finding, we derived a reduced search space of models that are likely to provide a good \ac{nnr} forecasting model on unseen data.

Future work concerns an extension of the current approach to include exogenous data.
Additionally, other artificial intelligence based methods, e.g., neural networks and support vector regression, are planned to be investigated.

\bibliographystyle{splncs}
\bibliography{databaseLiteraturePaper}

\end{document}